\documentclass[twocolumn,prl,amsmath,amssymb,showpacs]{revtex4}
\usepackage{graphicx}
\usepackage{dcolumn}
\usepackage{bm}
\usepackage{latexsym}
\usepackage{amstext}
\usepackage{amssymb}
\usepackage{amsxtra}

\newcommand{\bs}{\boldsymbol}
\newcommand{\op}{\omega_\perp}

\begin{document}
\title{Interference of an array of independent Bose-Einstein condensates}
\author{Zoran Hadzibabic, Sabine Stock, Baptiste Battelier, Vincent Bretin, and Jean Dalibard}
\affiliation{Laboratoire Kastler Brossel$^{*}$, 24 rue Lhomond,
75005 Paris, France}
\date{\today}

\begin{abstract}

We have observed high-contrast matter wave interference between 30
Bose-Einstein condensates with uncorrelated phases. Interference
patterns were observed after independent condensates were released
from a one-dimensional optical lattice and allowed to expand and
overlap. This initially surprising phenomenon is explained with a
simple theoretical model which generalizes the analysis of the
interference of two independent condensates.
\end{abstract}


\maketitle

Studies of Bose-Einstein condensates (BECs) loaded into the
periodic potential of an optical lattice have been continuously
growing in the recent years~\cite{bloc04physworld}. These systems
have a great potential for a range of applications such as
modelling of solid state
systems~\cite{jaks98mott,cata01,mors01,grei02mott,heck02},
preparation of low-dimensional quantum gases~\cite{pedr01,stof04},
tuning of atom properties~\cite{eier04,bloc04tonks}, trapped atom
interferometry~\cite{shin04}, and quantum information
processing~\cite{peil03,ott04}.

One of the most commonly used probes of these novel systems are
the interference patterns obtained when the gas is released from
the lattice, so that the wave packets emanating from different
lattice sites expand and overlap~\cite{orze01,grei02mott}. In
particular, the appearance of high-contrast interference fringes
in the resulting density distribution is commonly associated with
the presence of phase coherence between different lattice sites.

In this Letter, we study the interference of a regular array of
Bose-Einstein condensates with random relative phases. Independent
condensates, each containing $\sim 10^4$ atoms, are produced in a
one-dimensional (1D) optical lattice, in the regime where the rate
of tunnelling between the lattice sites is negligible on the time
scale of the experiment. Contrary to what could be naively
expected, we show that high-contrast interference fringes are
commonly observed in this system. We present a theoretical model
which quantitatively reproduces our experimental results, and show
that the periodicity of the lattice is sufficient for the
emergence of high-contrast interference patterns, even in the
absence of phase coherence between the condensates. This
conclusion is independent of the number of sites or the
dimensionality of the lattice. Our results generalize the analysis
of the interference between two independent
condensates~\cite{java96phas,nara96,cast97}. In the last part of
the paper we briefly discuss the potential of our system for
creating strongly number-squeezed states and report on an
unexplained heating effect which occurs for a narrow range of
lattice depths.

\begin{figure}
\centerline{\includegraphics[width=8.5cm]{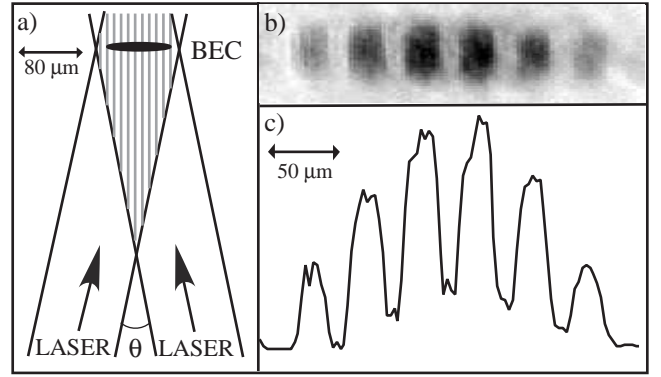}}
\vspace{-1mm}\caption{Interference of Bose-Einstein condensates
with uncorrelated phases. (a) A deep 1D optical lattice splits a
cigar-shaped condensate into 30 independent BECs. (b) Absorption
image of the cloud after 22~ms of expansion from the lattice. The
density distribution shows clear interference fringes of high
contrast. (c) Axial density profile of the cloud, radially
averaged over the central $25\,\mu$m. The fit described in the
text gives a fringe amplitude $A_1 =0.64$ for this image.}
\label{fig:fig1} \vspace{-4mm}
\end{figure}

Our experiments start with a quasipure $^{87}$Rb condensate with
$3\times10^5$ atoms in the $F=m_F=2$ hyperfine state. Condensates
are produced by radio-frequency evaporation in a cylindrically
symmetric magnetic trap. The harmonic trapping frequencies are
$\op/2\pi=74\,$Hz radially, and $\omega_z^{(0)}/2\pi=11\,$Hz
axially, leading to cigar-shaped condensates with a Thomas-Fermi
length of $L_{\rm TF}=84\,\mu$m, and a radius of $R_{\rm
TF}=6\,\mu$m.

We then ramp up the periodic potential created by a 1D optical
lattice. The lattice is superimposed on the magnetic trapping
potential along the long axis ($z$) of the cigar
(Fig~\ref{fig:fig1}a). Two equally polarized laser beams of
wavelength $\lambda=532$~nm intersect at an angle
$\theta=0.20$~rad to create a standing wave optical dipole
potential with a period of $d=\lambda/2~\sin(\theta/2)=2.7\,\mu$m.
The two beams are focused to waists of about $100\,\mu$m, and
carry a laser power of up to $220\,$mW each. The blue-detuned
laser light creates a repulsive potential for the atoms which
accumulate at the nodes of the standing wave, with the radial
confinement being provided by the magnetic potential. Along $z$,
the lattice potential has the shape
\begin{equation}
V(z)=\frac{V_{0}}{2}\cos(2\pi z/d)\ . \label{eq:potential}
\end{equation}
The lattice depth at full laser power is $V_0/h \approx 50\,$kHz,
and the number of occupied sites is $N = L_{\rm TF}/d \approx 30$.

Thanks to the long lattice period and the large hight of the
potential barrier between the sites, the 30 condensates can be
completely isolated from each other. The matrix element $J$ for
the tunnelling between neighboring sites scales as $E_{R}\,
e^{-2\sqrt{V_0/E_{R}}}$, where
$E_{R}=\hbar^{2}k^{2}/2m=h\times80$~Hz, $k=\pi/d$, and $m$ is the
atom mass. Our maximum lattice depth corresponds to $V_0 \approx
600\, E_{R}$, leading to a timescale $h/J$ of tens of years.

At full lattice depth, the gas at each site is in a quasi 2D
regime. The motion along $z$ is ``frozen out" because the
oscillation frequency $\omega_z/(2 \pi)$ is $4\,$kHz, while the
temperature and the chemical potential of the gas correspond to
frequencies smaller than $2.5\,$kHz. Each condensate is therefore
in the harmonic oscillator ground state along this direction, with
a density distribution given by a Gaussian of width $\ell
=\sqrt{\hbar/(2m\omega_z)}=120\,$nm$\,\ll d$.


%
\begin{figure}
\centerline{\includegraphics[width=8.5cm]{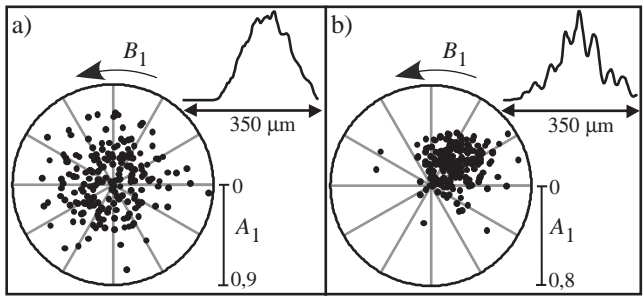}}
\caption{Polar plots of the fringe amplitudes and phases
$(A_1,B_1)$ for 200 images obtained under same experimental
conditions. (a) Phase-uncorrelated condensates. (b)
Phase-correlated condensates. Insets: Axial density profiles
averaged over the 200 images.} \label{fig:fig2} \vspace{-5mm}
\end{figure}

In a typical experiment, we ramp up the lattice to $600\, E_{R}$
in $\tau_{\rm ramp}=200\,$ms. After holding the atoms in the
lattice for $\tau_{\rm hold}=500\,$ms, the optical and magnetic
trapping potential are switched off simultaneously, and the
density distribution of the cloud is recorded by absorption
imaging after $t=22\,$ms of time-of-flight (ToF) expansion. The
hold time $\tau_{\rm hold}$ is sufficient for the phases of the
independently evolving condensates to completely decorrelate (see
below). Despite this, the images commonly show clear interference
fringes. A spectacular example of this surprising phenomenon is
given in Fig.~\ref{fig:fig1}b-c, where the contrast is $>60~\%$.
Note that high-contrast interference is also observed if the
lattice is ramped up before the end of the evaporation sequence,
so that the condensates are produced independently at different
lattice sites and ``have never seen one another"~\cite{ande86}.

We analyze the images by fitting the axial density profiles
(Fig.~\ref{fig:fig1}c) with $ [1+ A_1 \cos(B_1 + 2\pi
z/D)]\,G(z)$, where $G(z)$ is a Gaussian envelope. This procedure
extracts the first-harmonic modulation of the density
distribution, with the fitted fringe period of $39\,\mu$m in
agreement with the expected value $D=ht/(md)$. In
Fig.~\ref{fig:fig2}a, we summarize our results for 200
consecutively taken images. In most cases the fringe amplitude
$A_1$ is significant, with the mean value $\langle A_1 \rangle
=0.34$, and the standard deviation $\sigma_{A_1} = 0.17$. The
fringe phase $B_1$ is randomly distributed between 0 and $2\pi$.
Consequently, no periodic modulation remains visible if we average
the 200 density profiles.

In Fig.~\ref{fig:fig2}b, we contrast this with the interference of
30 phase-correlated condensates. Here, we ramp up the lattice in
$\tau_{\rm ramp}=3\,$ms and immediately release the BECs from the
trap, before their phases completely diffuse away from each other.
In this case, the fringe phases $B_1$ are clearly not randomly
distributed, and the sum of 200 images still shows interference
fringes of pronounced contrast. Phase correlations between the
separated condensates were lost if the lattice was left on for
$\tau_{\rm hold}\geq 3\,$ms.

We have numerically simulated our experiments with
phase-uncorrelated condensates using the following model. We
consider a 1D array of $N=30$ BECs initially localized at
positions $z_n=nd$ ($n=1,\ldots,N$), and assume that each
condensate is in a coherent state described by an amplitude
$\alpha_n$ and a phase
$\phi_n$~\cite{random_phases_footnote_fockvscoherent}. We set
$\alpha_n \propto n(N-n)$, corresponding to the Thomas-Fermi
profile of our BEC at the point when the lattice is switched on.
For expansion times $t \gg 1/\omega_z \approx 40\,\mu$s, the atom
density at position $z$ is given by~\cite{pedr01}:
 \begin{equation}
I(z) \propto \left| \sum_{n=1}^N \alpha_n\,
e^{i\phi_n}\;e^{im(z-z_n)^2/(2\hbar t)} \;e^{-(z-z_n)^2/Z_0^2}
 \right|^2
 \label{eq:Ix}
 \end{equation}
where $Z_0 = \hbar t/(m\ell)$. We neglected the effects of the
atomic interactions during the expansion. We perform a Monte-Carlo
analysis of $I(z)$ by assigning sets of random numbers to the
phases $\{\phi_n\}$. We convolve the resulting density profiles
with a Gaussian of $5\,\mu$m width to account for the finite
resolution of our imaging system, and then fit them in the same
way as the experimental data. The fitted fringe phase $B_1$ is
randomly distributed between 0 and $2\pi$. For the fringe
amplitude we find $\langle A_1\rangle^{\rm sim} =0.31$ and
$\sigma_{A_1}^{\rm sim} = 0.16$, in excellent agreement with the
experiment~\cite{unconvolved}.

In order to give a more intuitive explanation of our observations,
and generalize our analysis to arbitrarily large $N$, we now make
the following simplifications. We assume that all condensates
contain the same average number of atoms, $\alpha_n = \alpha$, and
that the expansion time $t$ is large, so that $\ell,z_n \ll
Z_0,\sqrt{\hbar t/m}$. In this case, Eq.~(\ref{eq:Ix}) can be
rewritten in the form $ I(z)\propto N \alpha^2 e^{-2z^2/Z_0^2}\,
F(z) $, where
\begin{eqnarray}
F(z) = 1+\sum_{n=1}^{N-1} A_n\, \cos(B_n + 2\pi nz/D)
 \label{eq:Fx}
 \end{eqnarray}
is a periodic function with period $D=ht/(md)$ and average value
1. $F(z)$ contains all the information about the contrast of the
interference pattern. The amplitude $A_n$ and the phase $B_n$ of
the $n$-th harmonic of $F(z)$ are given by the modulus and the
argument of $(2/N)\sum_{j=n+1}^Ne^{i(\phi_j-\phi_{j-n})}$.

If the $N$ condensates have the same phase, $F(z)$ corresponds to
the usual function describing the diffraction of a coherent wave
on a grating
\begin{equation}
F(z)=\frac{1}{N}\frac{\sin^2(N\pi z/D)}{\sin^2(\pi z/D)}\ .
\end{equation}
In this case, $F(z)$ has sharp peaks of hight $N$ and width $\sim
1/N$ at positions $z=pD$, where $p$ is an integer. In between the
peaks, $F(z)\sim 0$.

 For the case of uncorrelated phases $ \{ \phi_n \}$, using
Monte-Carlo sampling, we find for $10 \leq N \leq 10^4$:
\begin{equation}
\langle F_{\rm max}\rangle \sim 1.2\,\ln(N) \gg 1 \qquad \langle
F_{\rm min}\rangle \sim \frac{0.2}{N} \ll 1\
\end{equation}
where $F_{\rm max}$ and $F_{\rm min}$ are the maximum and the
minimum ``single-shot" values of $F(z)$, {\it i.e.} for a given
set $\{ \phi_n\}$. This shows that even for very large $N$, most
single-shot $F(z)$ are strongly modulated, with a contrast of
almost $100\%$. Since $F(z)$ is also periodic with period $D$,
each single-shot has the qualitative appearance of a high-contrast
interference pattern normally expected in the coherent case.
However, the exact shape of $F(z)$ and the position of its extrema
vary randomly from shot to shot.

The harmonic content of single-shot $F(z)$ is given by $C_n =
\langle A_n^2\rangle = 4(N-n)/N^2$, each $A_n$ resulting from
summing $N-n$ complex numbers with random phases and moduli $2/N$.
The large modulation of $F(z)$ for $N\to \infty$ can be understood
by noting that while the weight of each harmonic decreases, their
number increases~\cite{random_phases_footnote_toseeharmonics}. It
is interesting to contrast this with averaging $N$ shots of
two-condensate interference with an irreproducible
phase~\cite{andr97int}. In the latter case, the weight of the
first harmonic is similar ($\sim 1/\sqrt{N}$), but the absence of
higher harmonics leads to a vanishing contrast for large $N$.

We obtain similar results if we choose for the initial state a
Mott insulator with exactly one atom per lattice site (see
also~\cite{altm03}). In this case, the signal after ToF consists
of the set of coordinates $\zeta_1,\ldots,\zeta_N$ where the $N$
atoms are detected, and its harmonic content is given by:
\begin{equation}
C_n = \frac{4}{N(N-1)} \sum_{j=1}^N \sum_{j'\neq j} \left\langle
e^{i2\pi(\zeta_j-\zeta_{j'}) n/D}\right\rangle \nonumber
\end{equation}
where $\langle \ldots\rangle$ now denotes a quantum average. A
simple calculation gives $C_n = 4(N-n)/[N(N-1)]$.

All our arguments naturally extend into two and three dimensions.
In 3D, the function generalizing Eq.~(\ref{eq:Fx}) is $ F(\bs
r)=1+\sum_{\bs n} A_{\bs n} \cos( B_{\bs n} + 2\pi \bs n\cdot \bs
r/D) $ where $\bs n=(n_x,n_y,n_z)$ is a triplet of integers. This
is a periodic function in $x,y,z$ with a period $D$. A Monte-Carlo
analysis for a cubic lattice with $20\times 20\times 20$ sites
with random phases shows that $F(\bs r)$ is again strongly
modulated, with $\langle F_{\rm max}\rangle \simeq 12$ and
$\langle F_{\rm min}\rangle \ll 1$.

In the experimental study of the superfluid to Mott insulator
transition with cold atoms~\cite{grei02mott,stof04}, the
disappearance of interference fringes is observed in the
insulating domain, and used as one of the signatures for the loss
of long range phase coherence. This seems in contradiction with
our results. However one can reconcile the two findings by
noticing that the experiments~\cite{grei02mott,stof04} are
performed with a 3D system, observed by integrating the spatial
distribution $I(\bs r)$ along the line of observation $z$. In this
case, only the harmonics $(n_x, n_y, 0)$ are observed. In a 3D
experiment performed with $N_x \times N_y \times N_z$ sites, the
integration along $z$ thus reduces the modulation amplitude by
$\sqrt{N_z}$. In fact, the effect is similar to that of averaging
$N_z$ images in a 2D experiment, performed with $N_x\times N_y$
sites~\cite{random_phases_footnote_differentfromkeith}.

In the next part of this Letter, we briefly discuss the potential
of our setup for producing ``number-squeezed" states with large
occupation numbers~\cite{orze01}, where the atom number on each
site has a sub-Poissonian distribution with an average value $n_0
\gg 1$ and a standard deviation $\sigma < \sqrt{n_0}$. Such states
are important for atom interferometry and precision
measurements~\cite{bouy97}.

For simplicity, we restrict our discussion to the case of
translational invariance, without the quadratic magnetic potential
applied along the $z$-axis. In the ground state of the system, the
squeezing of the atom number on each site depends on the ratio of
the tunnelling rate $\tilde{t}=n_0 J/h$ and the effective strength
of the repulsive on-site interactions $U  \sim \mu/n_0$, where
$\mu$ is the chemical potential. To give a sense of scale, in our
case $n_0\sim 10^4$ and $U /h \sim 0.2\,$Hz. We can qualitatively
distinguish three regimes~\cite{orze01,zwer03}: (i) For $J \geq
n_0 \,U $, squeezing is negligible and the atom number on each
site follows a Poissonian distribution ($\sigma=\sqrt{n_0}\sim
100$). (ii) For $J \sim U $ ($ \tilde{t} \sim 200\,$Hz), squeezing
is significant and $\sigma$ is reduced to $n_0^{1/4} \sim 10$.
(iii) Finally, a phase transition to a Mott insulator state with
$\sigma<1$ occurs for $J \sim U /n_0$ ($\tilde{t} \sim 0.2\,$Hz).

In our setup, we can tune the value of $J$ across this full range,
and the  criterion for the Mott transition with $n_0 \sim 10^4$
atoms per site is satisfied for $V_0 \sim 100-150\,E_R$. However,
we point out two practical difficulties which arise for such large
values of $n_0$. First, for the system to be in its ground state,
all decoherence processes, such as particle loss, must have rates
lower than $\tilde{t}$. Second, the assumed translational
invariance must be insured to a sufficient level, so that the
potential energies at different lattice sites match to better than
$n_0 J=h\tilde{t}$. While it seems difficult to fulfill these
criteria for the values of $\tilde{t}$ low enough for the Mott
transition to occur, the regime of strong squeezing, $\sigma
\lesssim n_0^{1/4}$, should be accessible.

\begin{figure}
\centerline{\includegraphics[width=8.5cm]{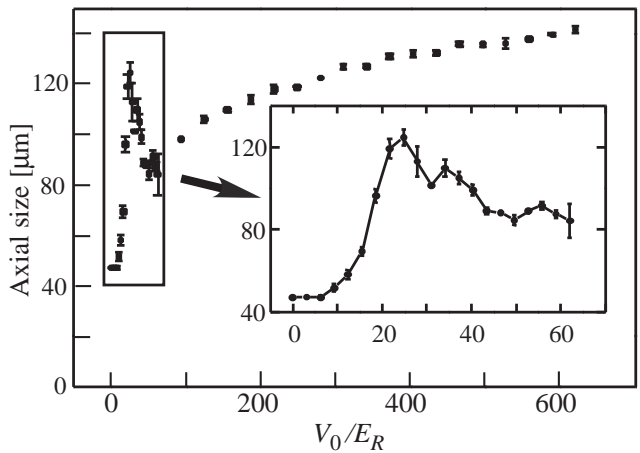}}
\vspace{-2mm}\caption{Axial size of the cloud after $22\,$ms of
expansion as a function of the lattice depth $V_0$. Inset: Zoom-in
on the region where an unexplained heating of the cloud occurs.
For all data points the lattice was raised in $\tau_{\rm
ramp}=200\,$ms and left on for $\tau_{\rm hold}=500\,$ms before
ToF. Error bars are statistical. Outside the heating region, the
slow increase of the axial size matches the expected dependence
$Z_0 \propto 1/\ell \propto V_0^{1/4}$.} \vspace{-5mm}
\label{fig:fig3}
\end{figure}

In exploring the range of lattice depths $0-150\,E_R$, we have
also observed an unexplained heating effect. In the region $20
<V_0/E_{R} <45$ we observe strong heating of the system which
peaks for $V_0/E_{R} \approx 25$ and 35 (Fig.~\ref{fig:fig3}). The
peak heating rate (measured from the {\it radial} size of the
cloud after ToF) is $\sim 200\,$nK/s as long as the gas is at
least partially condensed. However, once the condensate
disappears, heating of the thermal cloud becomes negligible ($<
10\,$nK/s). Outside $20 <V_0/E_{R} <45$, heating due to the
lattice is always negligible, independent of the lattice depth or
the condensed fraction. We could not attribute this heating to any
trivial technical effect. Also, the dynamical instability which
occurs at a finite relative velocity between the condensate and
the lattice (\cite{fall04} and refs. therein), should not be
relevant here.

In conclusion, we have studied a 1D periodic array of independent
condensates prepared in a deep optical lattice. We have shown that
most ``single-shot" realizations of this system show high-contrast
interference patterns in a time-of-flight expansion. We have
explained this effect with a simple model which naturally extends
to 3D lattices. This initially surprising result should be taken
into account in the ongoing studies of atomic superfluidity and
coherence in optical lattices, where the contrast of the
interference patterns is often used as a diagnostic.

\acknowledgments We thank A. Browaeys, T. Esslinger, A. Georges
and the ENS group for useful discussions. Z.H. acknowledges
support from a Chateaubriand grant, and S.S. from the
Studienstiftung des deutschen Volkes, DAAD, and the Research
Training Network "Cold Quantum Gases" HPRN-CT-2000-00125. This
work is supported by CNRS, Coll\`{e}ge de France, R\'egion Ile de
France, and DRED.


\begin{thebibliography}{30}
\expandafter\ifx\csname
natexlab\endcsname\relax\def\natexlab#1{#1}\fi
\expandafter\ifx\csname bibnamefont\endcsname\relax
  \def\bibnamefont#1{#1}\fi
\expandafter\ifx\csname bibfnamefont\endcsname\relax
  \def\bibfnamefont#1{#1}\fi
\expandafter\ifx\csname citenamefont\endcsname\relax
  \def\citenamefont#1{#1}\fi
\expandafter\ifx\csname url\endcsname\relax
  \def\url#1{\texttt{#1}}\fi
\expandafter\ifx\csname
urlprefix\endcsname\relax\def\urlprefix{URL }\fi
\providecommand{\bibinfo}[2]{#2}
\providecommand{\eprint}[2][]{\url{#2}}

\bibitem[*]{byline}
Unit\'e de Recherche de l'Ecole normale sup\'erieure et de
l'Universit\'e Pierre et Marie Curie, associ\'ee au CNRS.

\bibitem[{blo({\natexlab{a}})}]{bloc04physworld}
\bibinfo{note}{I. Bloch, Physics World, April 2004, pp. 25-29}.

\bibitem[{\citenamefont{Jaksch et~al.}(1998)\citenamefont{Jaksch, Bruder,
  Cirac, Gardiner, and Zoller}}]{jaks98mott}
\bibinfo{author}{\bibfnamefont{D.}~\bibnamefont{Jaksch}},
  \bibinfo{author}{\bibfnamefont{C.}~\bibnamefont{Bruder}},
  \bibinfo{author}{\bibfnamefont{J.~I.} \bibnamefont{Cirac}},
  \bibinfo{author}{\bibfnamefont{C.~W.} \bibnamefont{Gardiner}},
  \bibnamefont{and} \bibinfo{author}{\bibfnamefont{P.}~\bibnamefont{Zoller}},
  \bibinfo{journal}{Phys. Rev. Lett.} \textbf{\bibinfo{volume}{81}},
  \bibinfo{pages}{3108} (\bibinfo{year}{1998}).

\bibitem[{\citenamefont{Greiner et~al.}(2002)\citenamefont{Greiner, Mandel,
  Esslinger, H{\"a}nsch, and Bloch}}]{grei02mott}
\bibinfo{author}{\bibfnamefont{M.}~\bibnamefont{Greiner}},
  \bibinfo{author}{\bibfnamefont{O.}~\bibnamefont{Mandel}},
  \bibinfo{author}{\bibfnamefont{T.}~\bibnamefont{Esslinger}},
  \bibinfo{author}{\bibfnamefont{T.~W.} \bibnamefont{H{\"a}nsch}},
  \bibnamefont{and} \bibinfo{author}{\bibfnamefont{I.}~\bibnamefont{Bloch}},
  \bibinfo{journal}{Nature} \textbf{\bibinfo{volume}{415}}, \bibinfo{pages}{29}
  (\bibinfo{year}{2002}).

\bibitem[{\citenamefont{Hecker-Denschlag
  et~al.}(2002)\citenamefont{Hecker-Denschlag, Simsarian, H{\"a}ffner, McKenzie,
  Browaeys, Cho, Helmerson, Rolston, and Phillips}}]{heck02}
\bibinfo{author}{\bibfnamefont{J.}~\bibnamefont{Hecker-Denschlag}},
  \bibinfo{author}{\bibfnamefont{J.~E.} \bibnamefont{Simsarian}},
  \bibinfo{author}{\bibfnamefont{H.}~\bibnamefont{H{\"a}ffner}},
  \bibinfo{author}{\bibfnamefont{C.}~\bibnamefont{McKenzie}},
  \bibinfo{author}{\bibfnamefont{A.}~\bibnamefont{Browaeys}},
  \bibinfo{author}{\bibfnamefont{D.}~\bibnamefont{Cho}},
  \bibinfo{author}{\bibfnamefont{K.}~\bibnamefont{Helmerson}},
  \bibinfo{author}{\bibfnamefont{S.~L.} \bibnamefont{Rolston}},
  \bibnamefont{and} \bibinfo{author}{\bibfnamefont{W.~D.}
  \bibnamefont{Phillips}}, \bibinfo{journal}{J. Phys. B}
  \textbf{\bibinfo{volume}{35}}, \bibinfo{pages}{3095} (\bibinfo{year}{2002}).

\bibitem[{\citenamefont{Cataliotti et~al.}(2001)\citenamefont{Cataliotti,
  Burger, Fort, Maddaloni, Minardi, Trombettoni, Smerzi, and
  Inguscio}}]{cata01}
\bibinfo{author}{\bibfnamefont{F.~S.} \bibnamefont{Cataliotti}},
  \bibinfo{author}{\bibfnamefont{S.}~\bibnamefont{Burger}},
  \bibinfo{author}{\bibfnamefont{C.}~\bibnamefont{Fort}},
  \bibinfo{author}{\bibfnamefont{P.}~\bibnamefont{Maddaloni}},
  \bibinfo{author}{\bibfnamefont{F.}~\bibnamefont{Minardi}},
  \bibinfo{author}{\bibfnamefont{A.}~\bibnamefont{Trombettoni}},
  \bibinfo{author}{\bibfnamefont{A.}~\bibnamefont{Smerzi}}, \bibnamefont{and}
  \bibinfo{author}{\bibfnamefont{M.}~\bibnamefont{Inguscio}},
  \bibinfo{journal}{Science} \textbf{\bibinfo{volume}{293}},
  \bibinfo{pages}{843} (\bibinfo{year}{2001}).

\bibitem[{\citenamefont{Morsch et~al.}(2001)\citenamefont{Morsch, M{\"u}ller,
  Cristiani, Ciampini, and Arimondo}}]{mors01}
\bibinfo{author}{\bibfnamefont{O.}~\bibnamefont{Morsch}},
  \bibinfo{author}{\bibfnamefont{J.~H.} \bibnamefont{M{\"u}ller}},
  \bibinfo{author}{\bibfnamefont{M.}~\bibnamefont{Cristiani}},
  \bibinfo{author}{\bibfnamefont{D.}~\bibnamefont{Ciampini}}, \bibnamefont{and}
  \bibinfo{author}{\bibfnamefont{E.}~\bibnamefont{Arimondo}},
  \bibinfo{journal}{Phys. Rev. Lett.} \textbf{\bibinfo{volume}{87}},
  \bibinfo{pages}{140402} (\bibinfo{year}{2001}).

\bibitem[{\citenamefont{St{\"o}ferle et~al.}(2004)\citenamefont{St{\"o}ferle, Moritz,
  Schori, K{\"o}hl, and Esslinger}}]{stof04}
\bibinfo{author}{\bibfnamefont{T.}~\bibnamefont{St{\"o}ferle}},
  \bibinfo{author}{\bibfnamefont{H.}~\bibnamefont{Moritz}},
  \bibinfo{author}{\bibfnamefont{C.}~\bibnamefont{Schori}},
  \bibinfo{author}{\bibfnamefont{M.}~\bibnamefont{K{\"o}hl}}, \bibnamefont{and}
  \bibinfo{author}{\bibfnamefont{T.}~\bibnamefont{Esslinger}},
  \bibinfo{journal}{Phys. Rev. Lett.} \textbf{\bibinfo{volume}{92}},
  \bibinfo{pages}{130403} (\bibinfo{year}{2004});
\bibinfo{author}{\bibfnamefont{M.}~\bibnamefont{K{\"o}hl}},
  \bibinfo{author}{\bibfnamefont{H.}~\bibnamefont{Moritz}},
  \bibinfo{author}{\bibfnamefont{T.}~\bibnamefont{St{\"o}ferle}},
  \bibinfo{author}{\bibfnamefont{C.}~\bibnamefont{Schori}}, \bibnamefont{and}
  \bibinfo{author}{\bibfnamefont{T.}~\bibnamefont{Esslinger}},
  \bibinfo{journal}{cond-mat/0404338}  (\bibinfo{year}{2004}).

\bibitem[{\citenamefont{Pedri et~al.}(2001)\citenamefont{Pedri, Pitaevskii,
  Stringari, Fort, Burger, Cataliotti, Maddaloni, Minardi, and
  Inguscio}}]{pedr01}
\bibinfo{author}{\bibfnamefont{P.}~\bibnamefont{Pedri}},
  \bibinfo{author}{\bibfnamefont{L.}~\bibnamefont{Pitaevskii}},
  \bibinfo{author}{\bibfnamefont{S.}~\bibnamefont{Stringari}},
  \bibinfo{author}{\bibfnamefont{C.}~\bibnamefont{Fort}},
  \bibinfo{author}{\bibfnamefont{S.}~\bibnamefont{Burger}},
  \bibinfo{author}{\bibfnamefont{F.~S.} \bibnamefont{Cataliotti}},
  \bibinfo{author}{\bibfnamefont{P.}~\bibnamefont{Maddaloni}},
  \bibinfo{author}{\bibfnamefont{F.}~\bibnamefont{Minardi}}, \bibnamefont{and}
  \bibinfo{author}{\bibfnamefont{M.}~\bibnamefont{Inguscio}},
  \bibinfo{journal}{Phys. Rev. Lett.} \textbf{\bibinfo{volume}{87}},
  \bibinfo{pages}{220401} (\bibinfo{year}{2001}).

\bibitem[{\citenamefont{Eiermann et~al.}(2004)\citenamefont{Eiermann, Anker,
  M.Albiez, Taglieber, Treutlein, Marzlin, and Oberthaler}}]{eier04}
\bibinfo{author}{\bibfnamefont{B.}~\bibnamefont{Eiermann}},
  \bibinfo{author}{\bibfnamefont{T.}~\bibnamefont{Anker}},
  \bibinfo{author}{\bibnamefont{M.Albiez}},
  \bibinfo{author}{\bibfnamefont{M.}~\bibnamefont{Taglieber}},
  \bibinfo{author}{\bibfnamefont{P.}~\bibnamefont{Treutlein}},
  \bibinfo{author}{\bibfnamefont{K.~P.} \bibnamefont{Marzlin}},
  \bibnamefont{and} \bibinfo{author}{\bibfnamefont{M.~K.}
  \bibnamefont{Oberthaler}}, \bibinfo{journal}{cond-mat/0402178}
  (\bibinfo{year}{2004}).

\bibitem[{blo({\natexlab{b}})}]{bloc04tonks}
\bibinfo{note}{I. Bloch, private communication}.

\bibitem[{\citenamefont{Shin et~al.}(2004)\citenamefont{Shin, Saba, Pasquini,
  Ketterle, Pritchard, and Leanhardt}}]{shin04}
\bibinfo{author}{\bibfnamefont{Y.}~\bibnamefont{Shin}},
  \bibinfo{author}{\bibfnamefont{M.}~\bibnamefont{Saba}},
  \bibinfo{author}{\bibfnamefont{T.~A.} \bibnamefont{Pasquini}},
  \bibinfo{author}{\bibfnamefont{W.}~\bibnamefont{Ketterle}},
  \bibinfo{author}{\bibfnamefont{D.~E.} \bibnamefont{Pritchard}},
  \bibnamefont{and} \bibinfo{author}{\bibfnamefont{A.~E.}
  \bibnamefont{Leanhardt}}, \bibinfo{journal}{Phys. Rev. Lett.}
  \textbf{\bibinfo{volume}{92}}, \bibinfo{pages}{050405}
  (\bibinfo{year}{2004}).

\bibitem[{\citenamefont{Peil et~al.}(2003)\citenamefont{Peil, Porto, Tolra,
  Obrecht, King, Subbotin, Rolston, and Phillips}}]{peil03}
\bibinfo{author}{\bibfnamefont{S.}~\bibnamefont{Peil}},
  \bibinfo{author}{\bibfnamefont{J.~V.} \bibnamefont{Porto}},
  \bibinfo{author}{\bibfnamefont{B.~L.} \bibnamefont{Tolra}},
  \bibinfo{author}{\bibfnamefont{J.~M.} \bibnamefont{Obrecht}},
  \bibinfo{author}{\bibfnamefont{B.~E.} \bibnamefont{King}},
  \bibinfo{author}{\bibfnamefont{M.}~\bibnamefont{Subbotin}},
  \bibinfo{author}{\bibfnamefont{S.~L.} \bibnamefont{Rolston}},
  \bibnamefont{and} \bibinfo{author}{\bibfnamefont{W.~D.}
  \bibnamefont{Phillips}}, \bibinfo{journal}{Phys. Rev. A}
  \textbf{\bibinfo{volume}{67}}, \bibinfo{pages}{051603}
  (\bibinfo{year}{2003}).

\bibitem[{\citenamefont{Ott et~al.}(2004)\citenamefont{Ott, de~Mirandes,
  Ferlaino, Roati, T{\"u}rck, Modugno, and Inguscio}}]{ott04}
\bibinfo{author}{\bibfnamefont{H.}~\bibnamefont{Ott}},
  \bibinfo{author}{\bibfnamefont{E.}~\bibnamefont{de~Mirandes}},
  \bibinfo{author}{\bibfnamefont{F.}~\bibnamefont{Ferlaino}},
  \bibinfo{author}{\bibfnamefont{G.}~\bibnamefont{Roati}},
  \bibinfo{author}{\bibfnamefont{V.}~\bibnamefont{T{\"u}rck}},
  \bibinfo{author}{\bibfnamefont{G.}~\bibnamefont{Modugno}}, \bibnamefont{and}
  \bibinfo{author}{\bibfnamefont{M.}~\bibnamefont{Inguscio}},
  \bibinfo{journal}{cond-mat/0404201}  (\bibinfo{year}{2004}).

\bibitem[{\citenamefont{Orzel et~al.}(2001)\citenamefont{Orzel, Tuchman,
  Fenselau, Yasuda, and Kasevich}}]{orze01}
\bibinfo{author}{\bibfnamefont{C.}~\bibnamefont{Orzel}},
  \bibinfo{author}{\bibfnamefont{A.~K.} \bibnamefont{Tuchman}},
  \bibinfo{author}{\bibfnamefont{M.~L.} \bibnamefont{Fenselau}},
  \bibinfo{author}{\bibfnamefont{M.}~\bibnamefont{Yasuda}}, \bibnamefont{and}
  \bibinfo{author}{\bibfnamefont{M.~A.} \bibnamefont{Kasevich}},
  \bibinfo{journal}{Science} \textbf{\bibinfo{volume}{291}},
  \bibinfo{pages}{2386} (\bibinfo{year}{2001}).

\bibitem[{\citenamefont{Javanainen and Yoo}(1996)}]{java96phas}
\bibinfo{author}{\bibfnamefont{J.}~\bibnamefont{Javanainen}} \bibnamefont{and}
  \bibinfo{author}{\bibfnamefont{S.~M.} \bibnamefont{Yoo}},
  \bibinfo{journal}{Phys. Rev. Lett.} \textbf{\bibinfo{volume}{76}},
  \bibinfo{pages}{161} (\bibinfo{year}{1996}).

\bibitem[{\citenamefont{Naraschewski et~al.}(1996)\citenamefont{Naraschewski,
  Wallis, Schenzle, Cirac, and Zoller}}]{nara96}
\bibinfo{author}{\bibfnamefont{M.}~\bibnamefont{Naraschewski}},
  \bibinfo{author}{\bibfnamefont{H.}~\bibnamefont{Wallis}},
  \bibinfo{author}{\bibfnamefont{A.}~\bibnamefont{Schenzle}},
  \bibinfo{author}{\bibfnamefont{J.~I.} \bibnamefont{Cirac}}, \bibnamefont{and}
  \bibinfo{author}{\bibfnamefont{P.}~\bibnamefont{Zoller}},
  \bibinfo{journal}{Phys. Rev. A} \textbf{\bibinfo{volume}{54}},
  \bibinfo{pages}{2185} (\bibinfo{year}{1996}).

\bibitem[{\citenamefont{Castin and Dalibard}(1997)}]{cast97}
\bibinfo{author}{\bibfnamefont{Y.}~\bibnamefont{Castin}} \bibnamefont{and}
  \bibinfo{author}{\bibfnamefont{J.}~\bibnamefont{Dalibard}},
  \bibinfo{journal}{Phys. Rev. A} \textbf{\bibinfo{volume}{55}},
  \bibinfo{pages}{4330} (\bibinfo{year}{1997}).


\bibitem{ande86}
P.~W. Anderson, in {\it The Lesson of Quantum Theory}, edited by
J. de Boer, E. Dal, and O. Ulfbeck (Elsevier, Amsterdam, 1986).

\bibitem[{ran({\natexlab{a}})}]{random_phases_footnote_fockvscoherent}
\bibinfo{note}{This is equivalent to each condensate being in a Fock state,
  with the atom number following a Poisson distribution with a
  mean value $\alpha_n^2$ (see e.g.~\cite{cast97})}.

\bibitem{unconvolved}
The unconvolved, ``perfect resolution" profiles give $\langle A_1
\rangle^{\rm sim} = 0.43$ and $\sigma_{A_1}^{\rm sim} = 0.22$.


\bibitem[{ran({\natexlab{c}})}]{random_phases_footnote_toseeharmonics}
\bibinfo{note}{In practice, this means that a large modulation of $F(z)$ will
  be observed only if the imaging resolution is sufficient to detect a
  significant fraction of the harmonics.}

\bibitem[{\citenamefont{Andrews et~al.}(1997)\citenamefont{Andrews, Townsend,
  Miesner, Durfee, Kurn, and Ketterle}}]{andr97int}
\bibinfo{author}{\bibfnamefont{M.~R.} \bibnamefont{Andrews}},
  \bibinfo{author}{\bibfnamefont{C.~G.} \bibnamefont{Townsend}},
  \bibinfo{author}{\bibfnamefont{H.-J.} \bibnamefont{Miesner}},
  \bibinfo{author}{\bibfnamefont{D.~S.} \bibnamefont{Durfee}},
  \bibinfo{author}{\bibfnamefont{D.~M.} \bibnamefont{Kurn}}, \bibnamefont{and}
  \bibinfo{author}{\bibfnamefont{W.}~\bibnamefont{Ketterle}},
  \bibinfo{journal}{Science} \textbf{\bibinfo{volume}{275}},
  \bibinfo{pages}{637} (\bibinfo{year}{1997}).

\bibitem[{\citenamefont{Altman et~al.}(2003)\citenamefont{Altman, Demler, and
  Lukin}}]{altm03}
\bibinfo{author}{\bibfnamefont{E.}~\bibnamefont{Altman}},
  \bibinfo{author}{\bibfnamefont{E.}~\bibnamefont{Demler}}, \bibnamefont{and}
  \bibinfo{author}{\bibfnamefont{M.~D.} \bibnamefont{Lukin}},
  \bibinfo{journal}{cond-mat/0306226}  (\bibinfo{year}{2003}).

\bibitem[{ran({\natexlab{d}})}]{random_phases_footnote_differentfromkeith}
\bibinfo{note}{We also stress that the effects discussed here differ from the
  low-contrast interference which can arise in the Mott regime due to
  inhomogeneity of the system~\cite{kash02} or incomplete isolation between the
  lattice sites~\cite{roth03}}.

\bibitem[{\citenamefont{Kashurnikov et~al.}(2002)\citenamefont{Kashurnikov,
  Prokof'ev, and Svistunov}}]{kash02}
\bibinfo{author}{\bibfnamefont{V.~A.} \bibnamefont{Kashurnikov}},
  \bibinfo{author}{\bibfnamefont{N.~V.} \bibnamefont{Prokof'ev}},
  \bibnamefont{and} \bibinfo{author}{\bibfnamefont{B.~V.}
  \bibnamefont{Svistunov}}, \bibinfo{journal}{Phys. Rev. A}
  \textbf{\bibinfo{volume}{66}}, \bibinfo{pages}{031601(R)}
  (\bibinfo{year}{2002}).

\bibitem[{\citenamefont{Roth and Burnett}(2003)}]{roth03}
\bibinfo{author}{\bibfnamefont{R.}~\bibnamefont{Roth}} \bibnamefont{and}
  \bibinfo{author}{\bibfnamefont{K.}~\bibnamefont{Burnett}},
  \bibinfo{journal}{Phys. Rev. A} \textbf{\bibinfo{volume}{67}},
  \bibinfo{pages}{031602(R)} (\bibinfo{year}{2003}).

\bibitem[{\citenamefont{Bouyer and Kasevich}(1997)}]{bouy97}
\bibinfo{author}{\bibfnamefont{P.}~\bibnamefont{Bouyer}} \bibnamefont{and}
  \bibinfo{author}{\bibfnamefont{M.~A.} \bibnamefont{Kasevich}},
  \bibinfo{journal}{Phys. Rev. A} \textbf{\bibinfo{volume}{56}},
  \bibinfo{pages}{R1083} (\bibinfo{year}{1997}).

\bibitem[{\citenamefont{Zwerger}(2003)}]{zwer03}
\bibinfo{author}{\bibfnamefont{W.}~\bibnamefont{Zwerger}}, \bibinfo{journal}{J.
  Opt. B} \textbf{\bibinfo{volume}{5}}, \bibinfo{pages}{S9}
  (\bibinfo{year}{2003}).

\bibitem[{\citenamefont{Fallani et~al.}(2004)\citenamefont{Fallani, Sarlo, Lye,
  Modugno, Saers, Fort, and Inguscio}}]{fall04}
\bibinfo{author}{\bibfnamefont{L.}~\bibnamefont{Fallani}},
  \bibinfo{author}{\bibfnamefont{L.~D.} \bibnamefont{Sarlo}},
  \bibinfo{author}{\bibfnamefont{J.~E.} \bibnamefont{Lye}},
  \bibinfo{author}{\bibfnamefont{M.}~\bibnamefont{Modugno}},
  \bibinfo{author}{\bibfnamefont{R.}~\bibnamefont{Saers}},
  \bibinfo{author}{\bibfnamefont{C.}~\bibnamefont{Fort}}, \bibnamefont{and}
  \bibinfo{author}{\bibfnamefont{M.}~\bibnamefont{Inguscio}},
  \bibinfo{journal}{cond-mat/0404045}  (\bibinfo{year}{2004}).



\end{thebibliography}
\end{document}